\documentclass[12pt,epsf]{article}
\usepackage{graphicx}
\usepackage{epsfig}
\usepackage{amsmath}
\begin{document}
%%%%%%%%%%%%%%%%%%%%%%%%%%%%%%%%%%%%%%%%%%%

\def\a{\alpha}
\def\b{\beta}
\def\c{\varepsilon}
\def\d{\delta}
\def\e{\epsilon}
\def\f{\phi}
\def\g{\gamma}
\def\h{\theta}
\def\k{\kappa}
\def\l{\lambda}
\def\m{\mu}
\def\n{\nu}
\def\p{\psi}
\def\q{\partial}
\def\r{\rho}
\def\s{\sigma}
\def\t{\tau}
\def\u{\upsilon}
\def\v{\varphi}
\def\w{\omega}
\def\x{\xi}
\def\y{\eta}
\def\z{\zeta}
\def\D{\Delta}
\def\G{\Gamma}
\def\H{\Theta}
\def\L{\Lambda}
\def\F{\Phi}
\def\P{\Psi}
\def\S{\Sigma}

\def\o{\over}
\def\beq{\begin{eqnarray}}
\def\eeq{\end{eqnarray}}
\def\centeron#1#2{{\setbox0=\hbox{#1}\setbox1=\hbox{#2}\ifdim
\wd1>\wd0\kern.5\wd1\kern-.5\wd0\fi
\copy0\kern-.5\wd0\kern-.5\wd1\copy1\ifdim\wd0>\wd1
\kern.5\wd0\kern-.5\wd1\fi}}
\def\ltap{\;\centeron{\raise.35ex\hbox{$<$}}{\lower.65ex\hbox{$\sim$}}\;}
\def\gtap{\;\centeron{\raise.35ex\hbox{$>$}}{\lower.65ex\hbox{$\sim$}}\;}
\def\gsim{\mathrel{\gtap}}
\def\lsim{\mathrel{\ltap}}
\newcommand{\vev}[1]{ \left\langle {#1} \right\rangle }
\newcommand{\bra}[1]{ \langle {#1} | }
\newcommand{\ket}[1]{ | {#1} \rangle }
\newcommand{\EV}{ {\rm eV} }
\newcommand{\KEV}{ {\rm keV} }
\newcommand{\MEV}{ {\rm MeV} }
\newcommand{\GEV}{ {\rm GeV} }
\newcommand{\TEV}{ {\rm TeV} }
\def\diag{\mathop{\rm diag}\nolimits}
\def\Spin{\mathop{\rm Spin}}
\def\SO{\mathop{\rm SO}}
\def\O{\mathop{\rm O}}
\def\SU{\mathop{\rm SU}}
\def\U{\mathop{\rm U}}
\def\Sp{\mathop{\rm Sp}}
\def\SL{\mathop{\rm SL}}
\def\tr{\mathop{\rm tr}}

\def\IJMP{Int.~J.~Mod.~Phys. }
\def\MPL{Mod.~Phys.~Lett. }
\def\NP{Nucl.~Phys. }
\def\PL{Phys.~Lett. }
\def\PR{Phys.~Rev. }
\def\PRL{Phys.~Rev.~Lett. }
\def\PTP{Prog.~Theor.~Phys. }
\def\ZP{Z.~Phys. }

%%%%%%%%%%%%%%%%%%%%%%%%%%%%%%%%%%%%%%%%%%%%%%%%%%%%%%%%%%%%%%%%%%%%

\baselineskip 0.7cm

\begin{titlepage}

\begin{flushright}
UT-07-22
\end{flushright}

\vskip 1.35cm
\begin{center}
{\large \bf
    Composite Messenger Baryon as a Cold Dark Matter
}
\vskip 1.2cm
K. Hamaguchi, S. Shirai and T. T. Yanagida
\vskip 0.4cm

{\it  Department of Physics, University of Tokyo,\\
     Tokyo 113-0033, Japan}

\vskip 1.5cm

\abstract{Among various supersymmetric (SUSY) models, gauge-mediated SUSY breaking models with an ultra-light gravitino of mass $m_{3/2} \lsim {\cal O}(10)$ eV
 are very attractive, since they are completely free from notorious gravitino problems.
 A drawback of such a scenario is the absence of the supersymmetric cold dark matter.
 In this letter, we propose that a baryonic bound state of strongly interacting messenger particles, with a mass of ${\cal O}(100)$ TeV, can naturally be the cold dark matter. We also exemplify a model which realizes such a scenario.}
\end{center}
\end{titlepage}

\setcounter{page}{2}

\section{Introduction}

The presence of the dark matter (DM) in the universe is one of the most serious puzzles in particle physics 
and cosmology. If the dark matter is a stable particle $X$ and they were once in the thermal equilibrium 
at early stages of the universe, one may calculate its present energy density $\rho_X$ for a given  mass 
$m_X$ and for a given annihilation cross section $\sigma_{ann}$. If the annihilation processes take place 
via the standard-model gauge interactions, one finds $m_X\simeq {\cal O}(1)$ TeV to account for the present 
DM density. 
The well-known example for such a DM candidate is the lightest supersymmetry (SUSY) particle 
(LSP) in the SUSY standard model (SSM). On the contrary, if the $X$ is a strongly interacting particle 
and its annihilation cross section reaches the unitarity bound, one may conclude $m_X\simeq 100$ TeV~\cite{Griest:1989wd}.
It is very interesting to identify this mass scale with the SUSY-breaking scale $\Lambda_{SUSY}$ 
and consider the $X$ particle as a baryonic bound state of hidden quarks responsible for the dynamical SUSY
breaking~\cite{Dimopoulos:1996gy}.

In this letter, we propose a baryonic bound state of messenger particles in the gauge-mediation model 
\cite{GM} to be the observed dark matter $X$, as an alternative idea to~\cite{Dimopoulos:1996gy}. Here,
the mass of the dark matter $m_X$ and the SUSY-breaking 
scale $\Lambda_{SUSY}$ are ${\cal O}(100)$ TeV, which corresponds to the gravitino mass $m_{3/2}\simeq {\cal O}(10)$ 
eV. Notice that there is no cosmological gravitino problems for such a light gravitino 
\cite{light-gravitino} and hence the model is very safe and attractive.

\section{Composite Messenger Dark Matter}
In the minimal messenger model one often assumes a pair of messenger chiral multiplets, $\Psi^i +
{\bar \Psi}_i$ $(i=1-5)$, which transform as ${\bf 5 + 5^*}$ of the GUT gauge group SU(5)$_{\rm GUT}$, 
respectively.
In this case the messengers interact with themselves only through the SSM gauge interactions and their
annihilation cross section never reaches the unitarity bound. Thus, we extend the minimal model
so that the messengers have strong interactions~\cite{Izawa:2005yf} to enhance the annihilation cross section.
We consider that the above pair of messenger chiral multiplets, $\Psi^i + {\bar \Psi}_i$, 
belongs to nontrivial representations of a hidden strongly interacting gauge group G. We adopt
G=SU($N$) and assume the messengers to be fundamental and antifundamental representations of SU($N$), 
$\Psi^i_\alpha $ and ${\bar \Psi}_i^\alpha$, where $\alpha=1-N$. (The reason of this choice 
becomes clear below.) 
They form SU($N$)-singlet baryons and antibaryons, which are most likely stable because of 
a hidden baryon-number conservation. 

In general, 
the hidden baryons $B^{i_1 i_2\cdots i_N} = \epsilon^{\alpha_1\alpha_2\cdots\alpha_N} \Psi^{i_1}_{\alpha_1} \Psi^{i_2}_{\alpha_2} \cdots \Psi^{i_N}_{\alpha_N}$
 are charged under the SM gauge symmetries, and hence they cannot be the dark matter.\footnote{Charged dark matter 
 of mass ${\cal O}(100)$ TeV is excluded~\cite{Verkerk:1991jf}. 
Note that even the neutrino-like hidden baryon for $N=4$ ($B\sim D^c D^c D^c L$) is marginally excluded as the dominant component of DM for $m_B\simeq 100$ TeV, 
by the direct detection experiment (cf. \cite{Mack:2007xj}). 
Moreover, for $N=4$, the charged baryon $B\sim D^c D^c L L$ is likely to be lighter than the neutrino-like baryon for $m_{d}>m_{\ell}$.}
However, there is an exception.
Namely, for $N=5$, the hidden baryon
$B = \epsilon_{i_1 i_2\cdots i_5}\epsilon^{\alpha_1\alpha_2\cdots\alpha_5} \Psi^{i_1}_{\alpha_1} \Psi^{i_2}_{\alpha_2} \cdots \Psi^{i_N}_{\alpha_5}$
is completely neutral and hence can be the cold dark matter in the present universe. Therefore, we consider an $\SU(5)$ group as the strongly interacting hidden gauge group 
acting on the messenger multiplets.

Among the bosonic and fermionic composite baryons, the lightest one becomes the dark matter.\footnote{The other components decay into the lightest component by emitting SSM gauge multiplets.}
The hidden baryon $B=(\Psi)^5$ and anti-baryon ${\bar B}=({\bar \Psi})^5$ annihilate 
into hidden mesons
$M^j_i={\bar \Psi}^\alpha_i \Psi_\alpha^j$ via the hidden strong interactions. Thus, it is quite natural that the 
annihilation cross section reaches the unitarity bound. We may  naturally account for the observed
energy density of the DM for the hidden baryon mass $m_{B}\simeq 100$ TeV as explained above. 
The produced hidden mesons have SSM quantum numbers corresponding to $M\sim D^c \bar{D^c}$, $L \bar{L}$, $D^c \bar{L}$, and $L \bar{D^c}$,
 where we denote the messengers as  $\bar{\Psi}_i=(D^c, L)$ and 
$\Psi^i=(\bar{D^c}, \bar{L})$, suppressing the strong $\SU(5)$ index $\alpha$.
 For $M\sim L\bar{L}$ or $D^c\bar{D^c}$, they immediately decay into SSM gauge multiplets.
 The other mesons $M\sim D^c \bar{L}$ and $L \bar{D^c}$ can decay into SSM matter particles through superpotential,
$W= (1/M_{*})L {\bar D^c} H_u d^c$, where $H_u$ is the up-type Higgs multiplet and $d^c$ is the right-handed down quark in the SSM. $M_*$ is a high-scale mass such as the GUT scale and the Planck scale. We see that the decay of the hidden mesons can occur before the Big-Bang Nucleosynthesis starts ($t\sim 1$ sec) and thus causes no cosmological problem, as far as the meson masses are ${\cal O}(10)$ TeV.

\section{Example model}

In this section we show the existence of a gauge-mediation model which contains stable composite
baryons of messengers discussed in the previous section. However, we do not consider that the model 
presented in this section is the unique model for the realization of our proposal of the DM
candidate.

We consider the $D$-term gauge-mediation model proposed by Nakayama et al.\cite{Nakayama:2007cf}.
The model is based on the $\SU(N)\times U(1)_D$ gauge symmetry in the hidden sector. The SUSY is
supposed to be broken by the vacuum-expectation value of the $D$ term of the $U(1)_D$ gauge 
multiplet. This $U(1)_D$ gauge symmetry may be broken at the SUSY-breaking scale $\sqrt{D}$ \cite{Nakayama:2007cf} for the $D$-term generation.
The role of the $\SU(N)$ is to break the unwanted $R$ symmetry by hidden gaugino condensations.
We introduce messenger chiral multiplets to mediate both of the SUSY- and $R$-breaking effects
to the SSM sector. Thus, the messengers should carry both charges of $\SU(N)$ and $U(1)_D$
to generate SUSY-breaking masses for sfermions and gauginos in the SSM. Thus,
the messengers are necessarily strong interacting.
We introduce a pair of messenger chiral multiplets $\Psi$ and $\bar{\Psi}$ which are 
in the $({\boldsymbol N}, +1, {\bf 5})$ and $(\bar{\boldsymbol N}, -1, {\bf 5^*})$ 
representations of the $\SU(N) \times {\rm U(1)}_{D}\times{\rm SU(5)_{GUT}}$ 
gauge symmetry.

Then the integration of messengers induces the SSM gaugino masses as\footnote{The convention of the $D$-term is different from \cite{Nakayama:2007cf}.}
\begin{eqnarray}
M_{\rm gluino} &=&\kappa_1 \frac{N\alpha_3}{4\pi}\frac{ (g D)^4}{M_d^{7}},  \\
M_{\rm wino} &=&\kappa_1 \frac{N\alpha_2}{4\pi}\frac{(g D)^4 r^{10}}{M_d^{7}}, \\
M_{\rm bino} &=&\kappa_1 \frac{N\alpha_1}{4\pi}\frac{(g D)^4 }{M_d^{7}}\left( \frac{2}{5}+\frac{3}{5} r^{10} \right),
\end{eqnarray}
where $M_{d(\ell)}$ is the mass of the quark(lepton)-messenger $D^c (L)$, $r = M_d/M_{\ell}$, $g$  the ${\rm U(1)}_{D}$ coupling constant,
 $\alpha_1=5 \alpha _{\rm EM}/(3 \cos^2\theta_{W})$, and $\kappa_1$ a constant
\begin{equation}
\kappa_1 ={\cal O}(1)\times \frac{32 \pi^2 \Lambda_N^3}{M_d ^3},
\end{equation}
where $32 \pi^2\Lambda_N^3 $ is the condensation scale of the $\SU(N)$ gauginos. 
The scale $32 \pi^2\Lambda_N^3 $ is smaller than $M_{d,\ell}^3$ since the gaugino condensation 
takes place after the massive messengers are decoupled. As pointed out in \cite{Nakayama:2007cf}
we should take $32 \pi^2\Lambda_N^3 \simeq M_{d,\ell}^3$ to avoid the negative mass squared for 
sfermions. Thus we consider $\kappa_1$ to be an ${\cal O}(1)$ constant.

The SUSY-breaking masses for sfermions are given by
\begin{equation}
m^2_{\phi_i}=\kappa_2 N \frac{(g D)^4}{M_d^6} 2\left[
 \left(\frac{\alpha_3}{4\pi}\right)^2 C_3 (i) + 
 \left(\frac{\alpha_2}{4\pi}\right)^2 C_2 (i) r^6 + 
 \left(\frac{\alpha_1}{4\pi}\right)^2 \frac{3Y_i^2}{5} \left( \frac{2}{5}+\frac{3}{5} r^{6} \right) 
 \right],
\end{equation} 
where $\kappa_2$ is of order unity and $C_a(i)$ are Casimir invariants for the particle $\phi_i$.
As mentioned above, the perturbative calculation yields a negative $\kappa_2$ and hence we must
postulate that the $\SU(N)$ gauge interactions are strong at the messenger scale to avoid the 
negative mass squared. This requirement
is very important, since it naturally satisfies an independent requirement for the DM candidate
of mass ${\cal O}(100)$ TeV, that is, their annihilation cross section should be close to the unitarity 
bound. Namely, if the 
$\SU(N)$ interactions forming baryon and antibaryon bound states are strong at the 
messenger-mass scale, the baryon's annihilation cross section may naturally reach the 
unitarity bound, as like nucleons in QCD.

Let us discuss constraints on the above  parameters. $N$ should be 5 as discussed in the 
previous section to have neutral messenger baryon and antibaryon. We see that if $N$ is smaller 
than $6$  the perturbative GUT unification is 
maintained. Thus, we choose $N=5$ in the present analysis. Furthermore, the U(1)$_D$ gauge coupling 
$g$ is smaller than about $0.2$ at the messenger-mass scale ($\approx 100$ TeV) so that 
the ${\rm U(1)}_{D}$ gauge interactions are perturbative at the GUT scale. 
Moreover, $M_d> \sqrt{gD}$ must be satisfied, since otherwise scalar messengers become tachyonic and
the SM gauge symmetry is spontaneously broken.

We take $r=1$ at the GUT scale ($\approx 2\times 10^{16}$ GeV). Then, $M_d$ become heavier 
than $M_{\ell}$ below the GUT scale due to the renormalization effects. In fact, we find
$r\approx 2$ at the messenger-mass scale. However, the strong $\SU(5)$ interactions around the 
messenger scale make $r$ smaller and $r$ becomes down to 1.1 depending on how strong the $\SU(5)$
is. Finally, the gravitino mass is given by
\begin{equation}
m_{3/2}=\frac{1}{\sqrt{6}}\frac{D}{M_P},
\end{equation}
where $M_P = 2.4\times 10^{18}$ GeV is the reduced Planck scale. 

\begin{figure}[h!]
\begin{center}
\scalebox{1.39}{
\begin{picture}(220,150)(0,0)
\put(-27,0){\line(1,0){20}}
\put(-22,3){$\scriptstyle {\tilde G}_{3/2}$}

\put(5,11.4){\line(1,0){20}}
\put(10,12.4){$\scriptstyle h^0$}
\put(5,41){\line(1,0){20}}
\put(5,33){$\scriptstyle H^0\, A^0$}
\put(5,42){\line(1,0){20}}
\put(10,43.5){$\scriptstyle H^\pm$}

\put(37,11.7){\line(1,0){20}}
\put(42.,4.5){$\scriptstyle {\tilde{\chi}^0_1}$}
\put(37,13.3){\line(1,0){20}}
\put(42,16.5){$\scriptstyle {\tilde{\chi}^0_2}$}
\put(37,29.6){\line(1,0){20}}
\put(42.,33.5){$\scriptstyle {\tilde{\chi}^0_3}$}
\put(37,74.4){\line(1,0){20}}
\put(42,78.5){$\scriptstyle {\tilde{\chi}^0_4}$}

\put(69,12.9){\line(1,0){20}}
\put(74,15.5){$\scriptstyle {\tilde{\chi}^{\pm}_1}$}
\put(69,73.8){\line(1,0){20}}
\put(74,77.5){$\scriptstyle {\tilde{\chi}^{\pm}_2}$}

\put(101,87.5){\line(1,0){20}}
\put(106.,91.2){$\scriptstyle\tilde g$}

\put(133,131){\line(1,0){20}}
\put(133,132){\line(1,0){20}}
\put(133.2,133.2){$\scriptstyle\tilde d_L\,\tilde u_L$}
\put(133,122){\line(1,0){20}}
\put(133,123){\line(1,0){20}}
\put(133.2,114.){$\scriptstyle\tilde u_R\,\tilde d_R$}
\put(133,53.6){\line(1,0){20}}
\put(138.,48){$\scriptstyle\tilde e_L$}
\put(133,23.3){\line(1,0){20}}
\put(138.,26.2){$\scriptstyle\tilde e_R$}
\put(133,53){\line(1,0){20}}
\put(138.4,56){$\scriptstyle\tilde \nu_e$}

\put(165,114){\line(1,0){20}}
\put(170.,105){$\scriptstyle\tilde t_1$}
\put(165,127){\line(1,0){20}}
\put(165.,130.){$\scriptstyle\tilde t_2, \tilde b_2$}
\put(165,118){\line(1,0){20}}
\put(170.,119){$\scriptstyle\tilde b_1$}
\put(165,126){\line(1,0){20}}
\put(165,21.5){\line(1,0){20}}
\put(170.,23.5){$\scriptstyle\tilde \tau_1$}
\put(165,53.2){\line(1,0){20}}
\put(170.,56){$\scriptstyle\tilde \tau_2$}
\put(165,52.6){\line(1,0){20}}
\put(170.,46){$\scriptstyle\tilde \nu_{\tau}$}

\put(240,70){\small Mass }
\put(240,60){$\scriptstyle {\rm (GeV)}$ }
\put(220,-10){\vector(0,1){150}}
\put(220,0){\line(1,0){10}}
\put(232,-2){$\scriptstyle 0$}
\put(220,130){\line(1,0){5}}
\put(220,50){\line(1,0){10}}
\put(232,48.8){$\scriptstyle 500$}
\put(220,100){\line(1,0){10}}
\put(232,99.8){$\scriptstyle 1000$}
\put(220,10){\line(1,0){5}}
\put(220,20){\line(1,0){5}}
\put(220,30){\line(1,0){5}}
\put(220,40){\line(1,0){5}}
\put(220,50){\line(1,0){5}}
\put(220,60){\line(1,0){5}}
\put(220,70){\line(1,0){5}}
\put(220,80){\line(1,0){5}}
\put(220,90){\line(1,0){5}}
\put(220,100){\line(1,0){5}}
\put(220,110){\line(1,0){5}}
\put(220,120){\line(1,0){5}}
\end{picture}}
\caption[]{An example of mass spectrum. We set $D=310$ TeV, $M_{d}=32.6$ TeV, $g=0.01$, $\kappa_1 = 1$, $\kappa_2 = 3.5$, $r=1.1$, and $\tan\beta=38$. 
We have imposed that the B-term in the Higgs mass matrix is zero and $\mu=-130$GeV at the messenger scale.  Notice that the gluino is considerably lighter than the prediction of the standard gauge-mediation model \cite{GM}.
}
\label{fig:spectrum}
\end{center}
\end{figure}
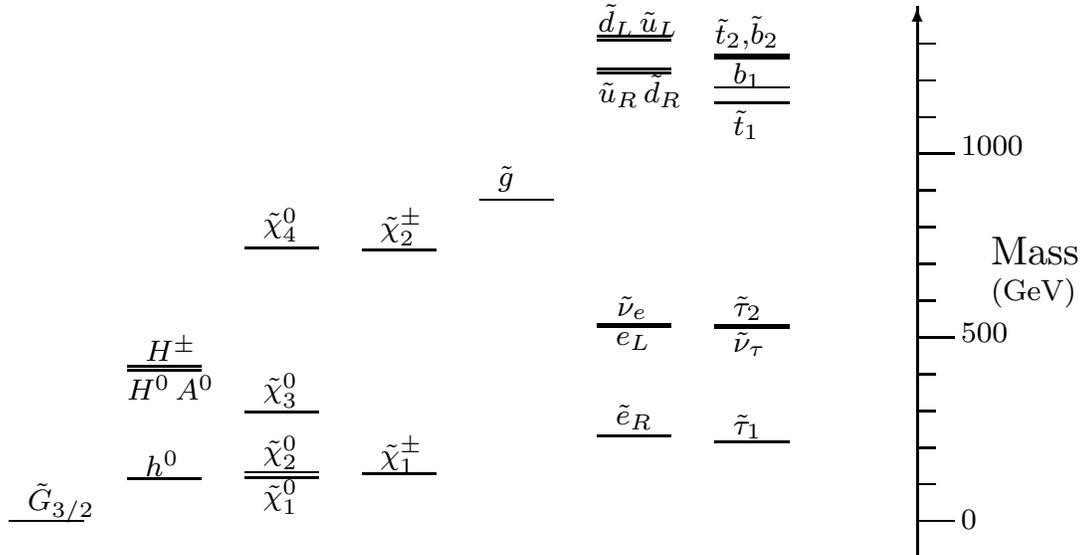

As an example, we take $m_{3/2}=16$ eV, which corresponds to $\sqrt{D} = 310$ TeV. The DM mass $m_B\simeq 100$ TeV implies the mass of messenger particles $M_d\simeq {\cal O}(10)$ TeV, and hence the constraint $M_d > \sqrt{gD}$ leads to $g \lsim {\cal O}(0.01)$. Here, we take  $M_d=32.6$ TeV and $g=0.01$ and show an example of mass spectrum in 
Fig.\ref{fig:spectrum}.
As for the parameter $r$, we have taken $r = 1.1$.
To evaluate the mass spectrum, we have used the program 
{\verb SOFTSUSY } 2.0.11~\cite{Allanach:2001kg}.

\section{Discussion}
The presence of the gravitino is an inevitable prediction of the supergravity. 
Cosmologically, it causes the notorious gravitino problems both in stable (LSP) and unstable (non-LSP) gravitino scenarios, 
which severely constrain the cosmological scenario and/or SSM models, depending on the gravitino mass.
 In that sense, the gravitino mass range $m_{3/2} < {\cal O}(10)$ eV is very attractive, since it is completely free from those problems.
 The only cosmological drawback of such a light gravitino scenario, compared with other cases, is the absence of the LSP dark matter, 
since the ${\cal O}(10)$ eV gravitino is too light to be the cold dark matter. 
In this Letter, we have suggested that there is indeed a natural dark matter candidate in such a low scale SUSY breaking scenario,
 namely, the composite messenger baryons of mass ${\cal O}(100)$ TeV.

Let us finally mention possible constraints and signatures. 
We should note that, although the constituent messenger particles of the composite DM baryons have the standard model gauge interactions, 
the confinement scale of the hidden $\SU(5)$ interaction ($\sim {\cal O}(100)$ TeV) is
 so large that the dark matter behaves essentially as a point-like SM singlet particle as far as the relevant energy scale is below ${\cal O}(100)$ TeV. Thus, the direct detection rate is negligible. Signals from the pair annihilation of DM in galactic center may become visible, but they are also likely to be buried in the background because of the small number density of the DM~\cite{Aharonian:2006wh}.

\end{document}